\documentclass[onecolumn]{aastex6}

\expandafter\ifx\csname natexlab\endcsname\relax\fi
\providecommand{\url}[1]{\href{#1}{#1}}
\providecommand{\dodoi}[1]{doi:~\href{http://doi.org/#1}{\nolinkurl{#1}}}
\providecommand{\doeprint}[1]{\href{http://ascl.net/#1}{\nolinkurl{http://ascl.net/#1}}}
\providecommand{\doarXiv}[1]{\href{https://arxiv.org/abs/#1}{\nolinkurl{https://arxiv.org/abs/#1}}}
\begin{document}

\title{Evolution of CME Properties in the Inner Heliosphere: Prediction for Solar Orbiter and Parker Solar Probe}

\author{Nada Al-Haddad\altaffilmark{1}}
\affil{IACS-Catholic University of America}

\author{No{\'e} Lugaz}
\affil{EOS-Department of Physics, University of New Hampshire}

\author{Stefaan Poedts}
\affil{Centre for mathematical Plasma Astrophysics (CmPA)- KU Leuven\\
Leuven, Belgium}

\author{Charles~J. Farrugia}
\affil{EOS-Department of Physics, University of New Hampshire}

\author{Teresa Nieves-Chinchilla\altaffilmark{2}}
\affil{NASA Goddard Space Flight Center}

\and

\author{Ilia~I. Roussev}
\affil{Centre for mathematical Plasma Astrophysics (CmPA)- KU Leuven}

\altaffiltext{1}{Work in this transcript was partially performed at KU Leuven}
\altaffiltext{2}{IACS-Catholic University of America}

\begin{abstract}
The evolution of the magnetic field and plasma quantities inside a coronal mass ejection (CME) with distance are known from statistical studies using data from 1~AU monitors, planetary missions, Helios, and Ulysses. This does not cover the innermost heliosphere, below 0.29~AU, where no data is publicly available  yet. Here, we describe the evolution of the properties of simulated CMEs in the inner heliosphere, using two different initiation mechanisms. We compare the radial evolution of these properties with that found from statistical studies based on observations in the inner heliosphere by Helios and MESSENGER. We find that the evolution of the radial size and magnetic field strength is nearly indistinguishable for twisted flux rope as compared to writhed CMEs. The evolution of these properties is also consistent with past studies, primarily with recent statistical studies using in-situ measurements and with studies using remote observations of CMEs.
 
\end{abstract}


\section{Introduction}

Coronal mass ejections (CMEs) occur at the Sun several times per day and impact Earth about once per week on average \citep[]{Yashiro:2004}. They carry magnetic energy, magnetic flux and plasma away from the Sun. In the past decade, there has been significant progress in determining CME properties in the corona from remote observations. The CME mass and speed are routinely determined from white-light coronagraph images \citep[]{Vourlidas:2010,Yashiro:2004}. Using the UltraViolet Coronagraph Spectrometer (UVCS) measurements, it is also possible to determine the full plasma properties using some assumptions for selected CMEs \citep[]{Bemporad:2010b}. Measuring magnetic fields in the corona is especially challenging, but recent observations with NSO/GONG have shown that it is possible to determine the magnetic field inside a CME up to 3~R$_\odot$ \citep[]{Dalmasse:2019, Tian:2013, Lin:1998}. A number of studies have shown that the magnetic field strength of a CME, as measured {\it in situ} at 1~AU is related to the reconnected magnetic flux during the eruption \citep[]{Qiu:2007, Kazachenko:2012, Moestl:2009b, Hu:2015}, as well as to the soft X-ray flux \citep[]{Moore:2001,Temmer:2008}. Nevertheless, all these observations only provide a few additional data points about the CME magnetic field in the lower corona. {\it In situ} measurements provide the only true measures of CME magnetic field and plasma quantities, however the majority of CMEs are only measured at one location in space \citep[]{Kilpua:2011}. Recent work based on hundreds of single-spacecraft measurements of CMEs near 1~AU \citep[]{Nieves:2018} or a subset of the simplest cases \citep[]{AlHaddad:2018} suggest that there might be multiple CME magnetic configurations \citep[]{Nieves:2019}. A key question is whether different magnetic configurations may be reflected in the way the CME properties evolve with distance. We aim at addressing this question here.

Measuring the same CME at different radial distances provides the best way to study the evolution of the CME properties during propagation. The seminal paper about magnetic clouds was in fact based on such a conjunction event between several spacecraft \citep[]{Burlaga:1981}. However, such multi-distance measurements of the same CME are rare, since there have not been many space missions in the inner heliosphere with the required instruments (magnetometers and plasma instruments), and, moreover, conjunctions are relatively rare. 

Even when two spacecraft measure the same CME at different radial distances, little can be learnt of the general decrease in the magnetic field inside CMEs with distance, or the expansion of CMEs from just two data points. Therefore, investigations of the evolution of CME properties in the inner heliosphere have focused on statistical studies. These assume that, if a sufficiently large  number of CMEs are observed at different radial distances, the radial dependence of the change in average CME properties  reflects the true average radial dependence of the CME properties. This assumption may not hold true if there are two or more populations of CMEs with different properties. For example, magnetic cloud and non-magnetic cloud CMEs may be two independent populations, instead of one population which differs based on how it is observed. In addition, CMEs with strong magnetic fields may expand faster than CMEs with low magnetic fields (due to the expansion being related to the total pressure imbalance). However, this assumption is the best that can be done with in situ measurements until a large fleet of space missions is sent in the inner heliosphere.

Excluding the recently launched Parker Solar Probe (PSP), from which data is not yet accessible to the public, Helios 1 and 2 are, to date, the two spacecraft which have gone closest to the Sun with a minimum distance of 0.29~AU, i.e.\ closer to the Sun than the orbit of Mercury, during the 1970s and 1980s. Most studies that have focused on the evolution of CME properties with distance have been based on Helios data in the inner heliosphere as well as Voyager and Ulysses data beyond the Earth's orbit \citep[]{Bothmer:1998,Liu:2005, Leitner:2007}. A few recent studies have used other planetary missions at Venus \citep[]{Jian:2008, Good:2016} and at Mercury \citep[]{Winslow:2015}. Lastly, with heliospheric remote observations enabled by the Solar Terrestrial Relations Observatory (STEREO), it has been possible to track the CME speed and radial width with distance in the inner heliosphere \citep[]{Savani:2009, Nieves:2012, Nieves:2013, Lugaz:2012b, Moestl:2014}, which have, overall, validated the results found from the statistical studies. The main findings of these studies are summarized when compared to the findings from the numerical simulations in the next section. Some numerical studies have focused on the evolution of the CME speed in simulations \citep[e.g., see][]{Jacobs:2007}, but few have investigated the evolution of the CME radial extent and magnetic field strength as we do here. The rest of the article is organized as follows. In section \ref{sec:model}, we give a quick overview of the simulations used in the present study. In section \ref{sec:properties}, we investigate the change of the CME properties with distance in the inner heliosphere and compare with recent studies. Lastly, in section \ref{sec:discussion}, we discuss our results and conclude. 

\begin{figure}[t]
\centering
{\includegraphics[height=7cm,width=7cm]{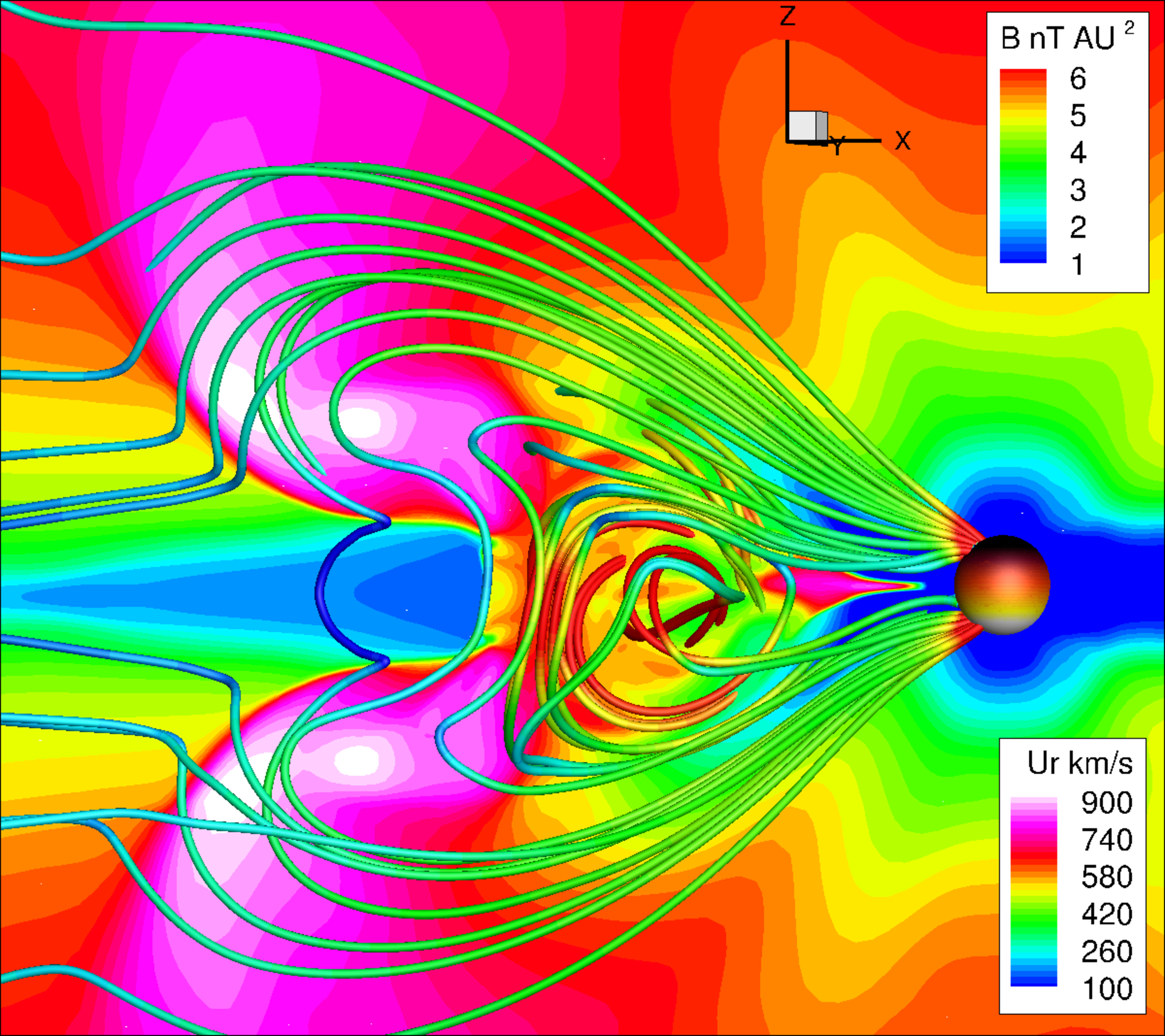}} 
{\includegraphics[height=7cm,width=7cm]{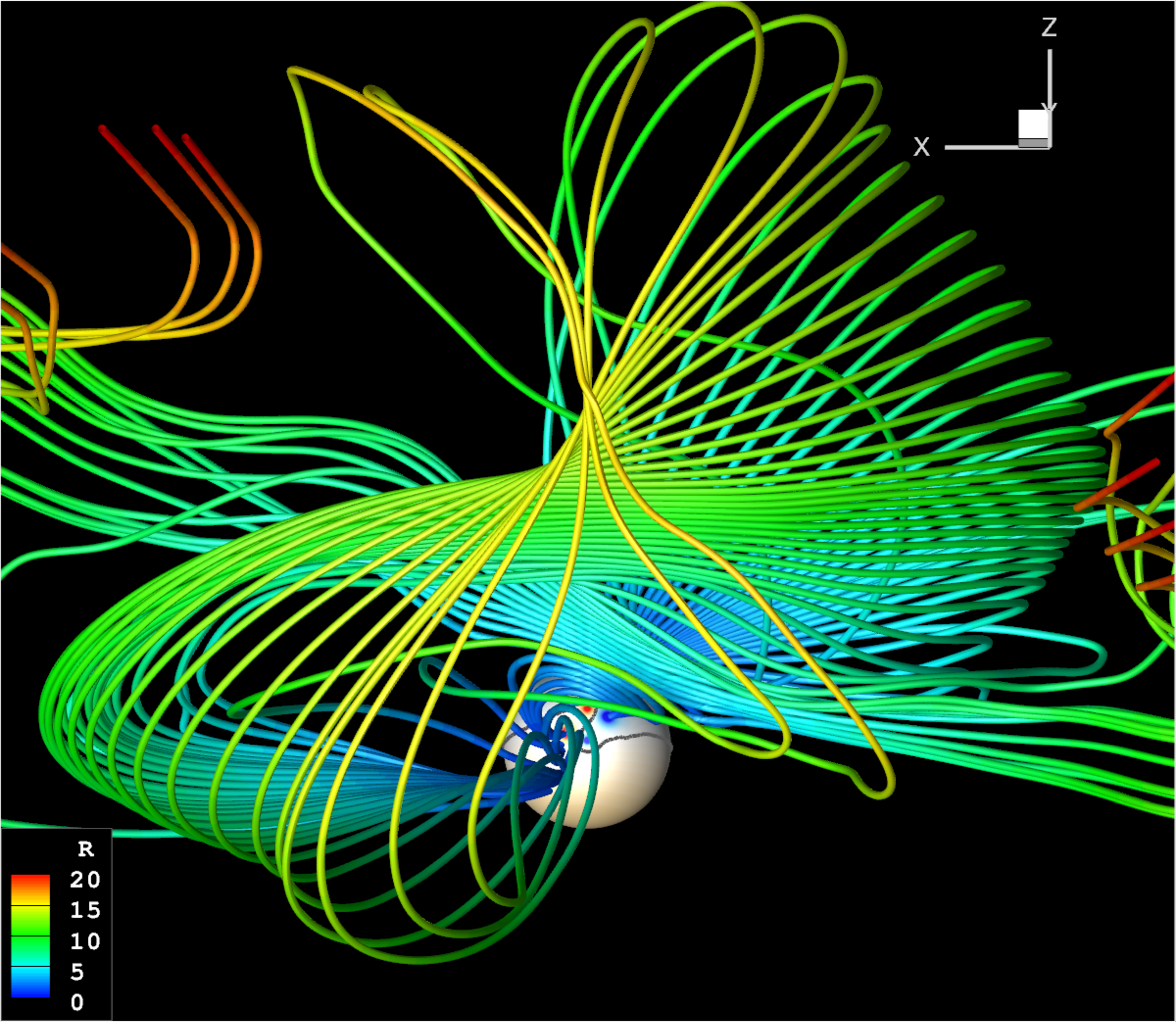}} 
\caption{Three-dimensional magnetic field lines of the two simulated CMEs used in this work as they reach the upper corona. Left: Twisted magnetic flux rope three hours after its initiation from \citet{Lugaz:2014}. The color shows the radial velocity. Right: Writhed magnetic ejecta four hours after its initiation from \citet{AlHaddad:2011}. The color shows the radial distance of the magnetic field lines.}
\label{fig:sims}
\end{figure}

\section{Numerical Models} \label{sec:model}
\begin{figure}[t]
\figurenum{2}
\centering
{\includegraphics[width=8.1cm]{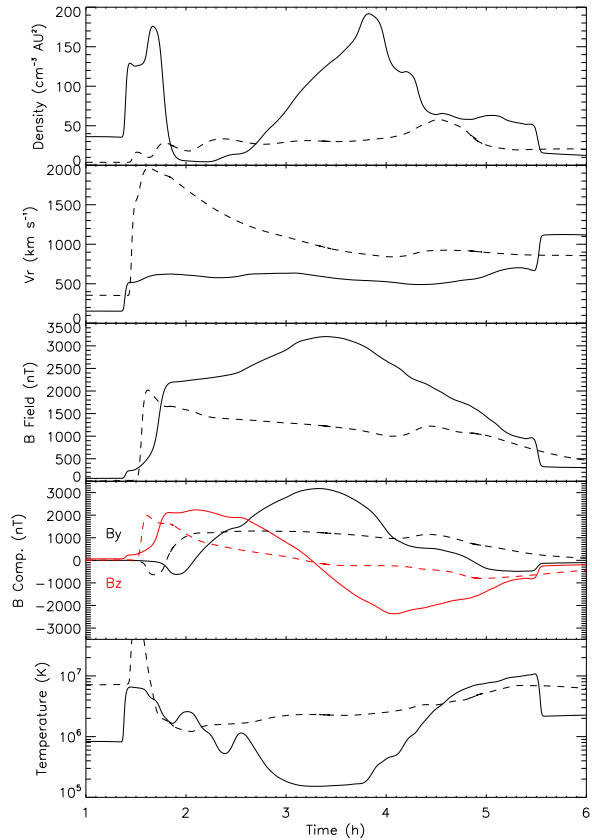}}
\caption{Simulated {\it in-situ} measurements for the twisted flux rope CME (solid) and the writhed CME (dashed) at 15~$R_\odot$ in the solar equatorial plane and along the central meridian. The panels show from top to bottom, the density scaled to its value at 1~AU, the radial velocity, the total magnetic field strength, the $B_y$ (black) and $B_z$  (red) components of the magnetic field vector and the plasma temperature. The twisted flux rope CME is shifted forward by 2.5 hours to match the shock arrival times at the simulated spacecraft. Figure adapted from \citet{ AlHaddad:2019}.}
\label{fig:sat}
\end{figure}

The simulations to create the synthetic spacecraft measurements are both performed with BATSRUS \citep[]{Powell:1999}, part of the Space Weather Modeling Framework \citep[SWMF,][]{Toth:2005,Toth:2012}. They do not correspond to real events but to idealized and simplified situations. This selection arose from the need to study as simple a CME as possible, without the influence of equatorial coronal holes or complex solar wind structures. The two simulations are performed with different models but the resulting CMEs have somewhat similar orientations, as described below. This eliminates the differences in CME evolution known to occur for CMEs of different orientation with respect to the interplanetary magnetic field (IMF) \citep[]{Chane:2006}. In the following subsections, we give the important details for both simulations. The {\it in situ} measurements for the writhed simulation have been analyzed in \citet{AlHaddad:2011}, and the two sets of {\it in situ} measurements have been compared in detail in \citet{AlHaddad:2019}. Figures~\ref{fig:sims} and \ref{fig:sat} show the 3-D view of the magnetic field line of the two simulated CMEs in the upper corona and synthetic spacecraft measurements at 15 $R_\odot$, respectively. 

\subsection{Twisted Flux Rope Simulation}
The simulation to create the synthetic satellite measurements for the twisted flux rope CME is the same one as the isolated case published in \citet{Lugaz:2014} that we previously analyzed in detail in \citet{ AlHaddad:2019}.The solar magnetic field is set up as non-tilted dipole with an octupole component with a maximum field strength at the solar surface of 5.5~G.  The solar wind model is that of \citet{Holst:2010} for a single fluid without heat conduction. These parameters reproduce solar minimum conditions. The CME is initiated with an out-of-equilibrium flux rope \citep[]{Gibson:1998}. The same model has been used to create simulated spacecraft measurements, which compare morphologically very well with typical CME structures measured at 1~AU \citep[]{Manchester:2004b}. The CME corresponds to a moderately fast one, with a speed of 700~km~s$^{-1}$ at 20~$R_\odot$ and a strong magnetic field of $\sim 3000$~nT at 15~$R_\odot$ (see solid lines in Figure~\ref{fig:sat}). Once propagated to 1~AU, the maximum magnetic field is about 60~nT ($-28$~nT for the southward $B_z$ component). The CME is initiated at the solar equator along the central meridian with an orientation parallel to the solar equator. It is a north-east-south (NES) type flux rope following the convention of \citet{Mulligan:1998} and \citet{Bothmer:1998}. The evolution of the CME during the first 9 hours of its propagation (up to 0.14~AU) was described in details in \citet{Lugaz:2013b}. In addition, it was found that its radial expansion was comparable to theoretical and statistical works \citep[]{Gulisano:2010}. 

\subsection{Writhed CME Simulation}
The simulation to create the satellite measurements for the writhed CME is the same one as the quadrupolar case that we previously analyzed and described in \citet{AlHaddad:2011, AlHaddad:2019}. To summarize, the solar magnetic field is set up with a global dipole as well as a quadrupolar active region which is positioned at the solar equator and centered at the central meridian. The maximum field strength at the poles is $\pm 4$~G and the active region field is $\sim 70$~G. The simulation includes a bimodal solar wind characteristics of solar minimum, which is created using the model of \citet{Roussev:2003b}. The CME is initiated with the shearing method described in more details in  \citet{Roussev:2007} and \citet{Jacobs:2009}. The inner magnetic spots in the quadrupolar active regions are sheared, and shearing flows are imposed onto the solar surface. The maximum shearing speed is about 3\% of the local Alfv{\'e}nic speed or $\sim 90$~km~$^{-1}$ and lasts 30 minutes.This method produces an eruption with minimal twist, but with a succession of writhed field lines (see right panel of Figure~\ref{fig:sims}). The simulated CME reaches a speed close to 2000~km~s$^{-1}$ at 20$~R_\odot$ and a north-to-south rotation of the magnetic field vector. The maximum magnetic field strength along the CME nose at 15~$R_\odot$ is about 2000~nT (see dashed lines in Figure~\ref{fig:sat}).


\section{Evolution of CME Properties in the Inner Heliosphere}\label{sec:properties}

\begin{figure}[ht!]
\figurenum{3}
\plottwo{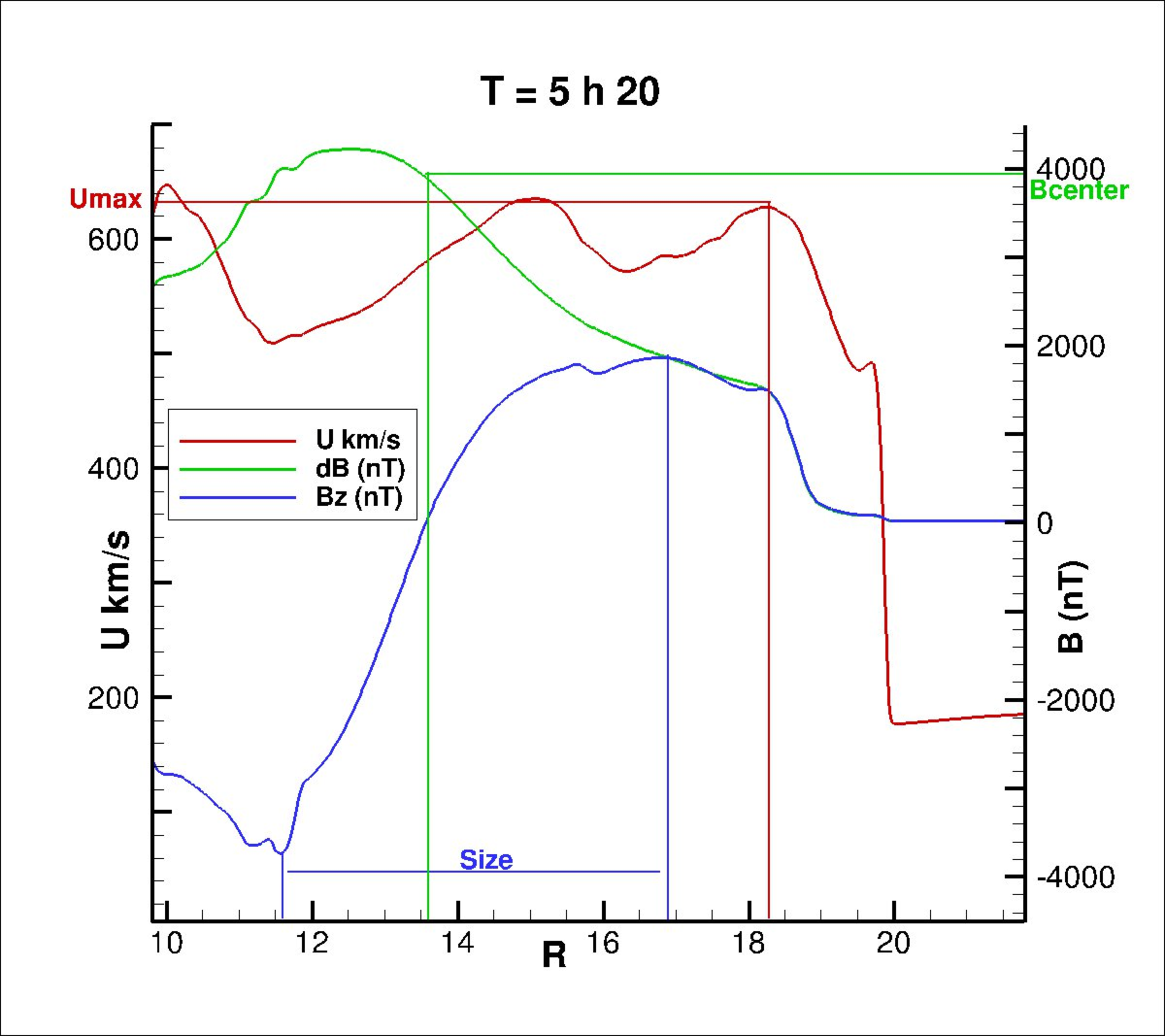}{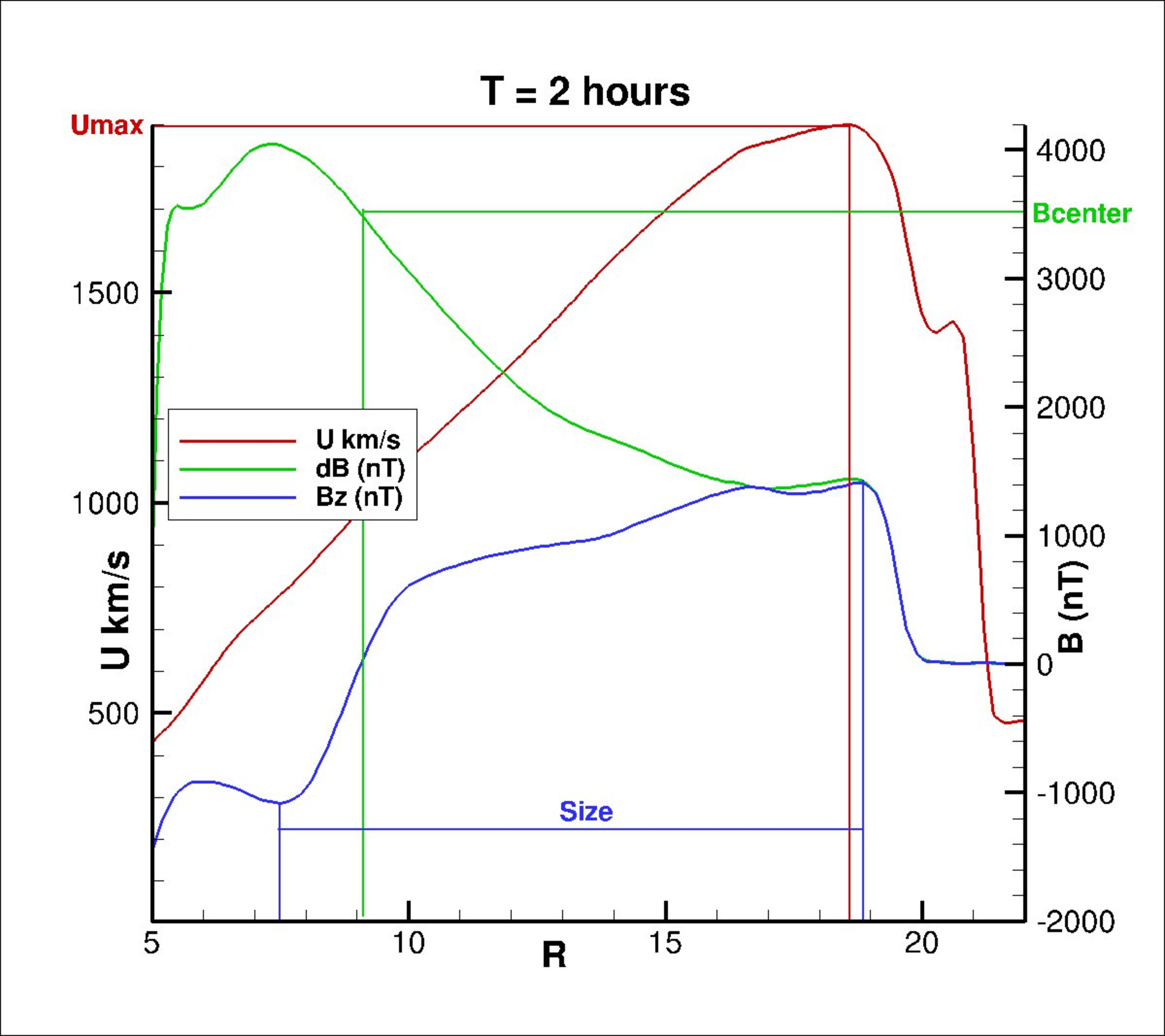}
\caption{Procedure to extract the magnetic field at the center of the CME, the CME size, and the CME maximum speed. The two panels show cuts through the twisted CME simulation (left) and writhed CME simulation (right) at approximately the same distance (shock close to 20 R$_\odot$). The line plots show the velocity (red), magnetic field increase from its steady-state value (green) and $B_z$ component of the magnetic field (blue) plotted {\it vs.} radial distance from the center of the Sun. The CME size is calculated from $B_z$ minimum (vertical blue line to the left) to maximum (vertical blue line to the right), the maximum speed is calculated inside the CME (value: horizontal red line), and the magnetic field at the center of the cloud is the magnetic field at the location where $B_z = 0$ (location: vertical green line, value: horizontal green line). See text for details.}

\label{fig:example}
\end{figure}

As described above, in most past studies, satellite measurements at different heliocentric distances are used to determine the evolution of the CME properties in a statistical manner. Through numerical simulations, the entire 3-D structure of the CME is known at every time step. However, in order to compare with past studies, it is important to estimate the simulated CME properties in a way comparable to the procedure executed on real satellite data. When possible, it is also important to improve upon these measurements in a way not possible with real data. For this reason, we made 1-D cuts every 20-30 minutes in the simulation along the $x =0, z=0$ line (the $y$-axis), i.e.\ right through the CME nose along the propagation direction. These 1-D cuts then represent ``snapshots'' of the variation of the CME magnetic field and plasma quantities along the radial distance at a given time. The 1-D cuts are then analyzed in a way similar to what would be done for satellite measurements except that the quantities are plotted with respect to the radial distance instead of time. Figure~\ref{fig:example} shows two examples of these 1-D cuts, and highlights how we measured the quantities plotted in the following section for the twisted CME and writhed CME simulation, at approximately the same radial distance from the Sun. 

The main difference between this approach and real satellite measurements is that the simulation provides a ``snapshot'' at a given time of the CME, whereas satellite measurements provide a view at a given location of the passage of the CME. For many of the CME properties, a snapshot is, in fact, a superior view, since it allows to remove the effects of the CME evolution during the passage through the measuring spacecraft. For example, many studies (see below) have investigated the evolution of the CME radial size based on satellite measurements. However, because the CME continues to expand and to be deformed and eroded during its passage over the spacecraft (see review by \citet[]{Manchester:2017} for example), the size calculated from spacecraft measurements is in fact an average over the duration of the CME crossing (which is typically of the order of one day, at 1~AU). 1-D cuts from simulations will provide the actual CME size at a given time. More discussion of the effect of CME expansion can be found in \citet[]{Demoulin:2008}.

\subsection{Evolution of the Magnetic Field at the Center of the CME}

\begin{figure}[t!]
\figurenum{4}
\plottwo{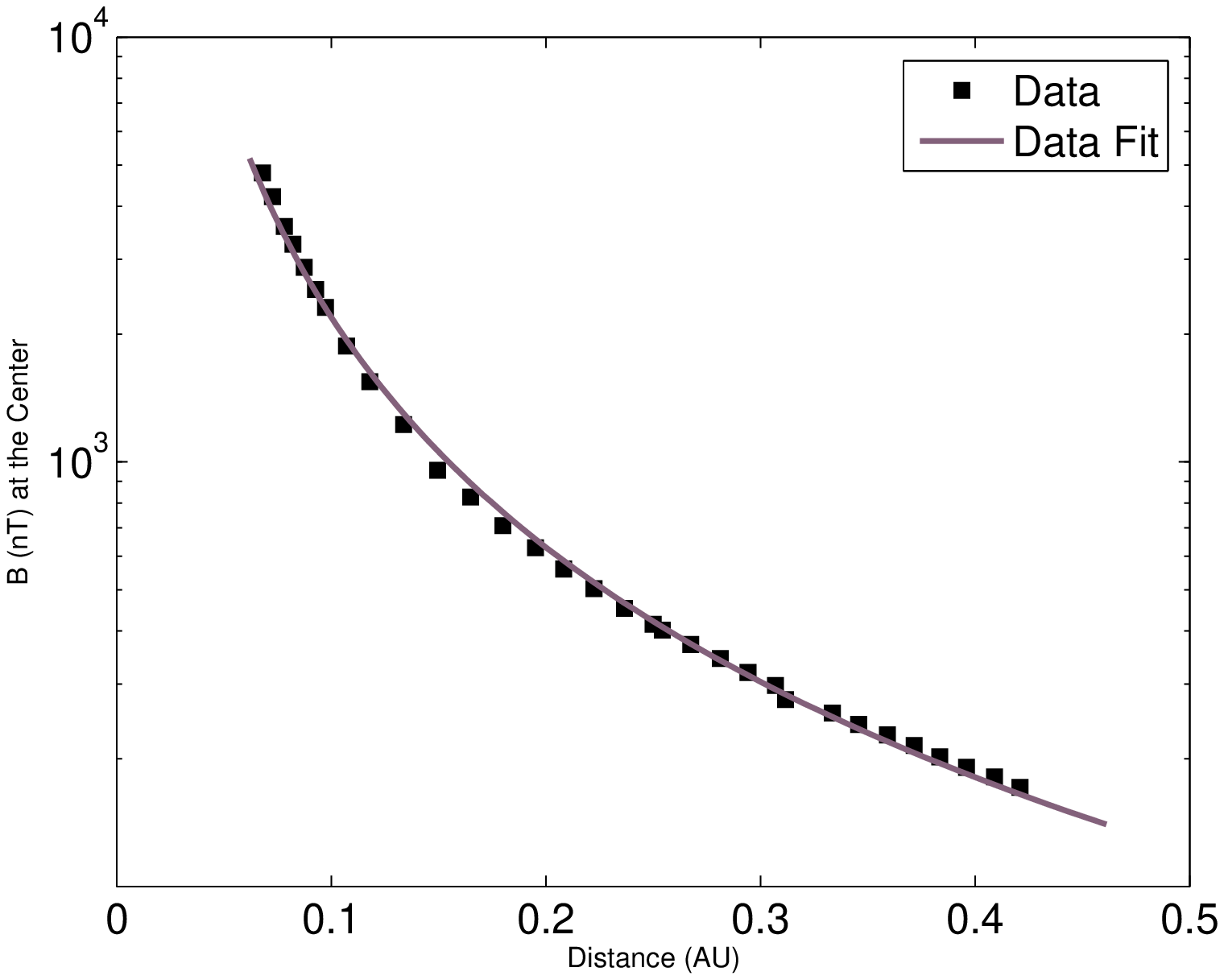}{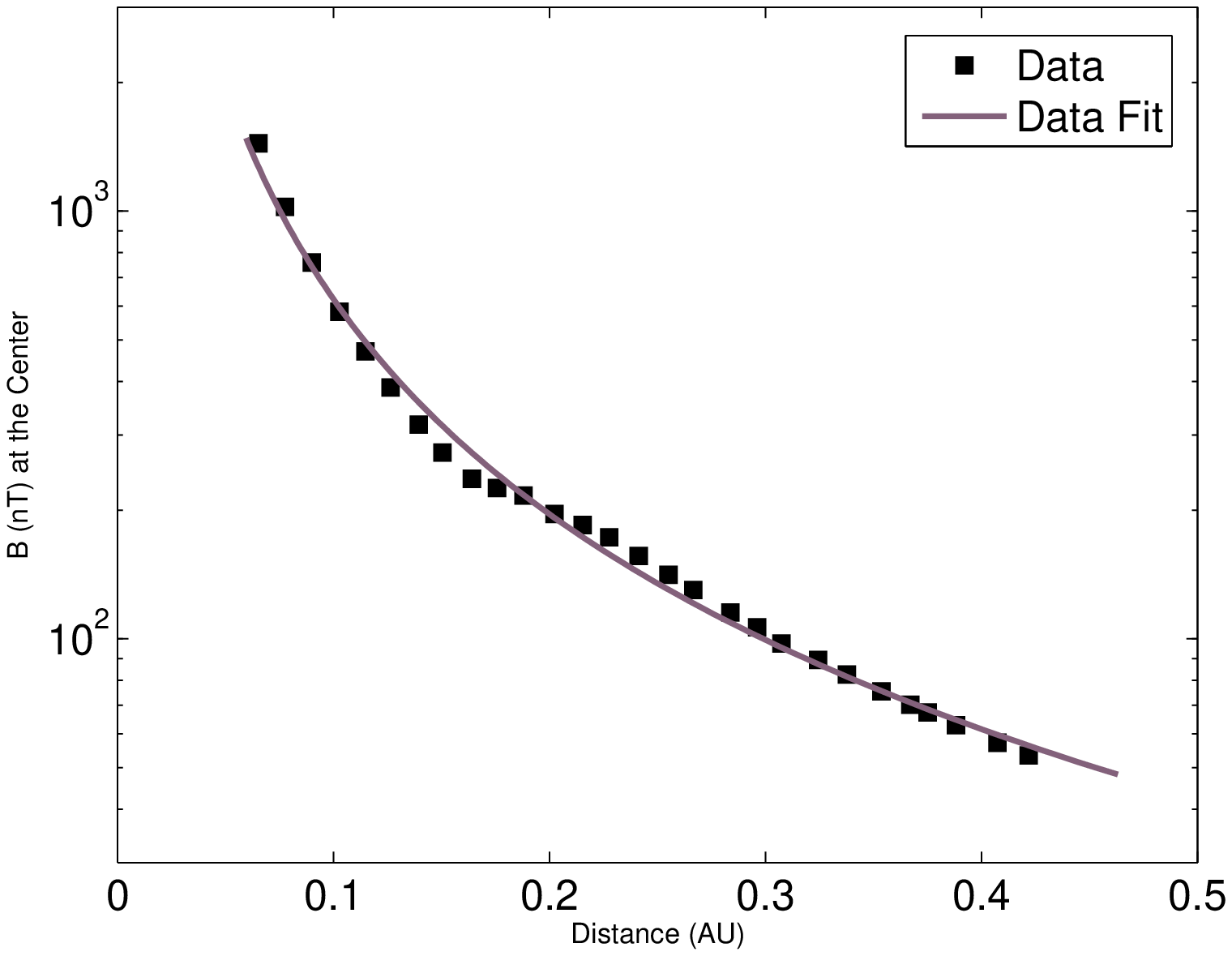}
\caption{Magnetic field at the center of the CME for the twisted (left) and writhed (right) simulations. The purple lines show power-law fits to the data.}
\label{fig:Bfield}
\end{figure}

We first track the center of the CME, which we define as the location where $B_z = 0$ (green line in Figure~\ref{fig:example}). For both CMEs, the poloidal field is in the $(x,z)$ plane; since the cut is made along $x=0$, the location where $B_z = 0$ allows us to track the center of the CME. The same definition was used, for example, in \citet{Lugaz:2005b}. Figure~\ref{fig:Bfield} shows the evolution of the magnetic field in the CME center from the solar corona to Mercury's orbit ($\sim 0.4$~AU). By fitting these two curves with a power law, we find that: 
\begin{eqnarray*}
B_\mathrm{twist} &=&  34.8\, R^{-1.8},\\
B_\mathrm{writhe} &=& 13.3\, R^{-1.67},
\end{eqnarray*}
where $B$ (nT) and $R$ (AU) are the magnetic field strength at the CME center and the distance of the CME center, respectively.
The R-squared for the fits are 0.998 and 0.994, respectively. 

These magnetic field strength at the center vs. distance relations can be compared to the statistical relations found by \citet{Gulisano:2010} of $B = 10.9 R^{-1.85}$, by \citet{Liu:2005} of $B = 7.4 R^{-1.4}$, by \citet{Leitner:2007} of $B = 17.7 R^{-1.73}$, all of which use Helios data, and recent results by \citet{Winslow:2015} of $B = 12.2 R^{-1.89 \pm 0.14}$ using STEREO and MESSENGER data. Theoretical considerations by \citet{Demoulin:2009b} found that $B$ is expected to vary with distance as $R^{-1.4}$, somewhat on the lower end of these values.

For both simulations, the radial distance dependence is very close to these published results, especially with that of \citet{Gulisano:2010} and \citet{Winslow:2015} and \citet{Leitner:2007}. The fact that the trend from our simulations is very close to results from these past studies is somewhat unexpected, since the Helios data used in the study by \citet{Gulisano:2010} and \citet{Leitner:2007} is from solar cycle 21, the MESSENGER data used in the study by \citet{Winslow:2015} from solar cycle 24 and our results are based on two numerical simulations. From the values of the magnetic field at 1~AU, the writhe CME has a magnetic field strength of a typical CME, whereas that for the twist CME is rather large, but comparable to that of very strong CMEs measured a few times per solar cycle, and significantly lower than that of extreme CMEs, such as the 2012 July 23 CME where the peak field was $\sim$ 109 nT.

Several conclusions can be drawn from these results. First, the writhed and twisted simulations are almost indistinguishable, based on the radial dependence of the magnetic field strength at the center of the CME. Note that the magnetic fields strength for the twisted CME decreases faster with distance, however,  both simulations are within the range of observed values. Second, both  simulations are consistent with statistical results using different missions and data from different solar cycles. Third, as compared to these statistical results, the simulations (i) include data points between the Sun and 0.3~AU, and (ii) provide the evolution for a single CME, confirming the statistical results based on $\sim$100 CMEs measured at different radial distances.

\subsection{Evolution of the CME Radial Extent}

\begin{figure}[ht!]
\figurenum{5}
\plottwo{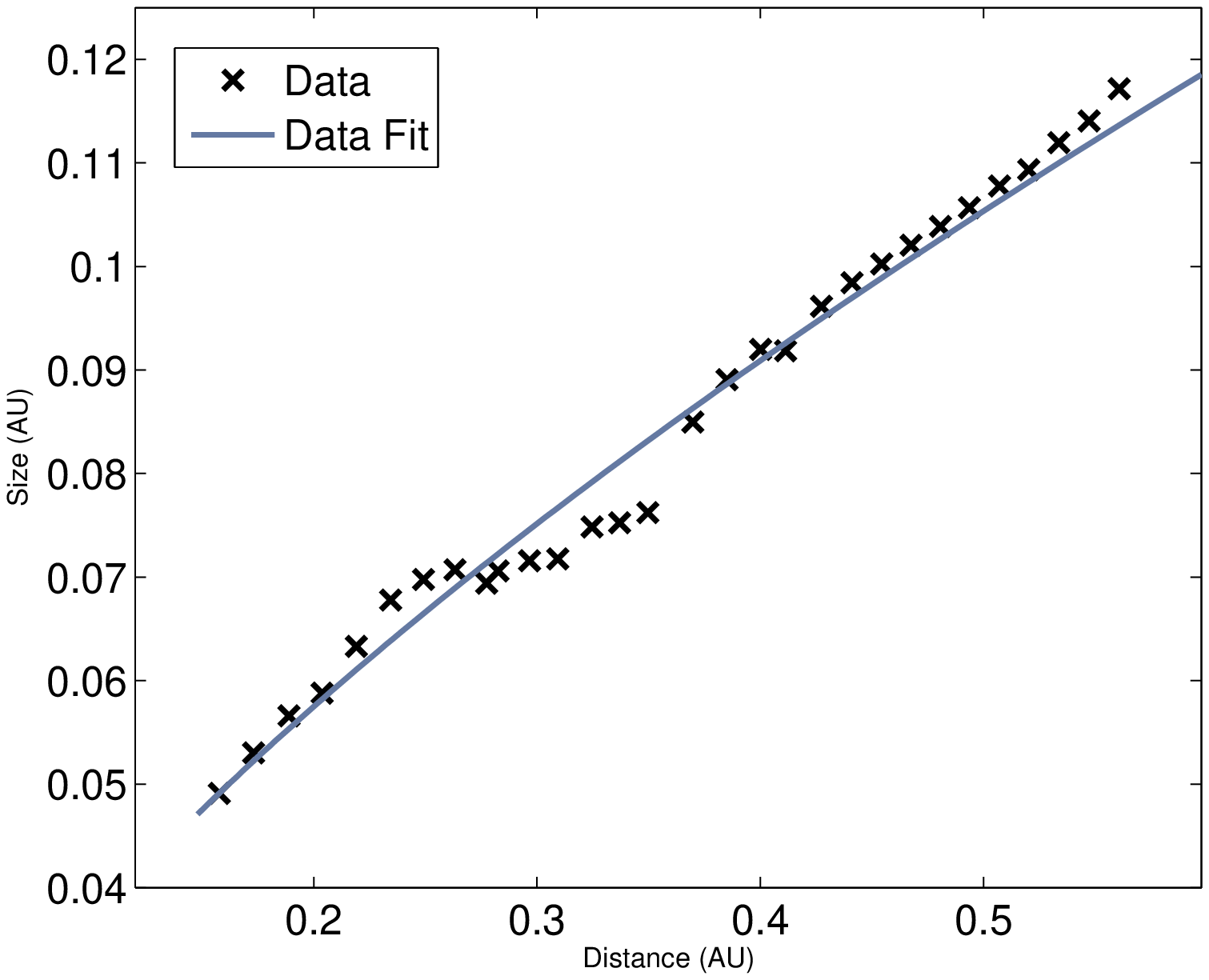}{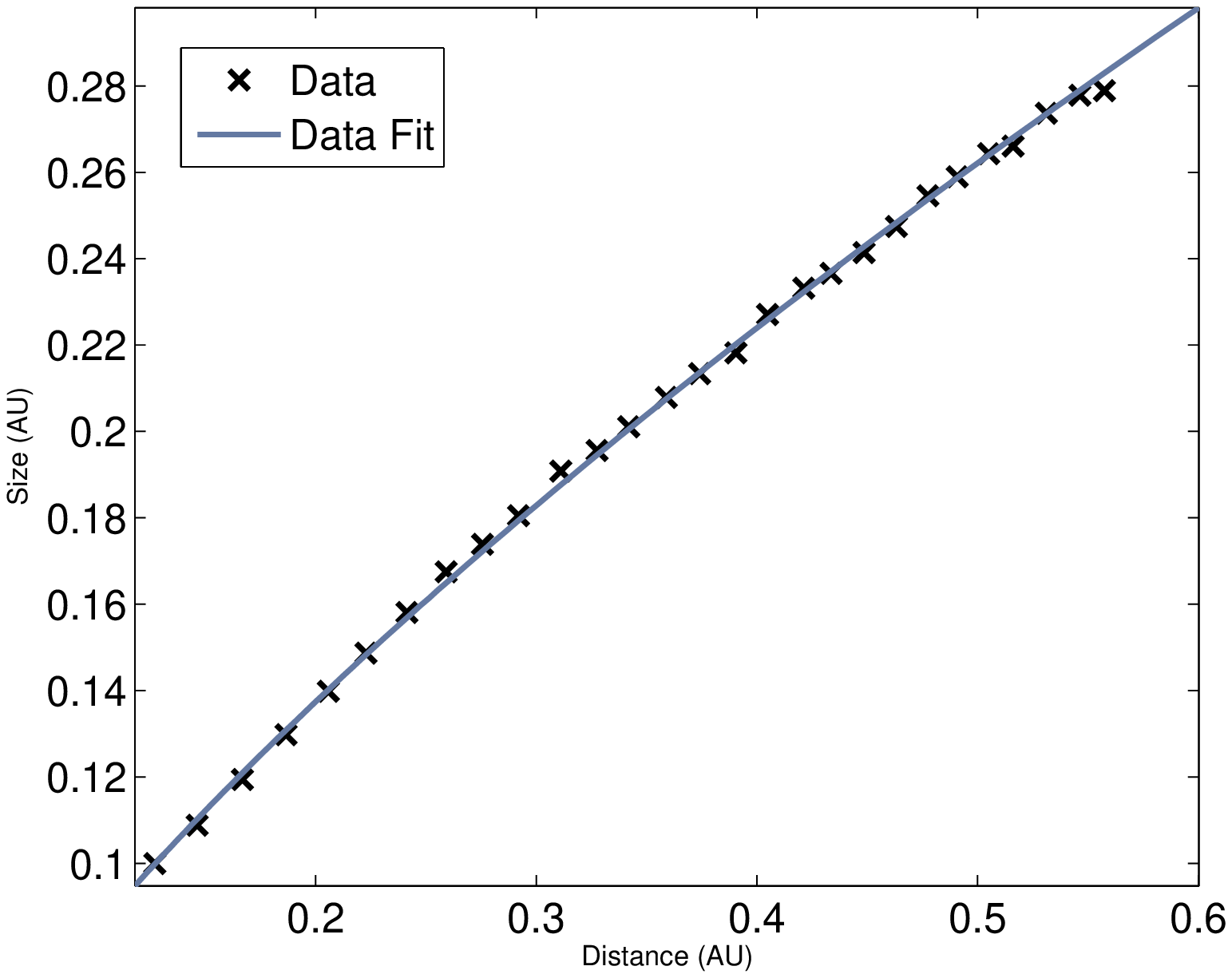}
\caption{Radial size of the CME for the twisted (left) and writhed (right) simulations as plotted versus the distance of the CME front. The blue lines are power-law fits to the data}
\label{fig:size}
\end{figure}

In order to measure the CME radial extent, we track the location of the maximum and minimum in $B_z$ (blue line in Figure~\ref{fig:example}). We calculate the radial dependency of this extent with the position of the front of the CME, since this approximately corresponds to the size of the CME when it ``hits'' a spacecraft at that location. Note that there could be other criteria to determine the CME size (ratio of poloidal to axial fields, threshold of the magnetic field strength larger than a given percentage above the background, etc.). Our choice of using the minimum and maximum of $B_z$ to determine the CME size results in a size smaller than that obtained using other techniques (approximately by 20--30\%). However, we have found that this technique yields a robust estimate of the CME size which can be easily followed during the simulation. In our experience, this is not the case of the size estimates using other criteria. 
Figure~\ref{fig:size} shows the evolution of the CME radial extent from the solar corona to 0.5~AU. By fitting these two curves, using least squares, we find that: 
\begin{eqnarray*}
S_\mathrm{twist} &=&  0.166\, R^{0.661},\\
S_\mathrm{writhe} &=& 0.427\, R^{0.705},
\end{eqnarray*}
where $S$(AU) and $R$(AU) are the CME radial extent and the distance of the CME front, respectively. The R-squared values are 0.979 and 0.999 respectively. 
Once again, this can be compared with past studies: 
\begin{eqnarray*}
S_\mathrm{Bothmer} &=& 0.24\, R^{0.78},\\
S_\mathrm{Gulisano} &=& 0.23\, R^{0.78},\\
S_\mathrm{Leitner} &=& 0.20\, R^{0.61},\\
S_\mathrm{Liu} &=&  0.25\, R^{0.92},\\
S_\mathrm{Lugaz} &=& 0.21\, R^{0.82},\\
S_\mathrm{Nieves-Chinchilla} &=& 0.2\, R^{0.74},\\
S_\mathrm{Savani} &=& 0.27\, R^{0.65}.
\end{eqnarray*}
%
%
The first four results are from previously cited studies based on the statistical analyses of in-situ measurements with Helios \citep[including the original study by][]{Bothmer:1998}. The last three studies are case studies based on remote-sensing observations with STEREO/SECCHI, investigating one CME each by \citet{Lugaz:2012b}, \citet{Nieves:2013} and \citet{Savani:2009}, respectively. Comparison with the newer results based on remote observations is somewhat more appropriate since: (1) these results are based on one CME only; (2) they are based on the size of the CME at a given time (that of the image) and not an average size over the spacecraft crossing; and (3) they cover the inner-most heliosphere where remote observations have less error (typically within 0.4~AU, as for our fits). 

As for the magnetic field at the center of the CME, the radial dependency of the CME radial extent is in good agreement with past studies, and it appears to have the best agreement with studies based on remote observations \citep[]{Savani:2009, Nieves:2013}, as anticipated for the three reasons given above. The expected radial size of the CME at 1~AU is on the low end for the twisted flux rope CME and on the high end for the writhed CME. However, these variations are well within the range observed at 1~AU. \citet{Lepping:2006} found a radial extent for magnetic clouds of 0.249 $\pm 0.122$~AU (1-$\sigma$ value) with about 10\% of the events (11 out of 114) with a radial extent larger than 0.4~AU (largest extent of 0.724 for the December 14, 2006 CME). In contrast to the evolution of the magnetic field, the evolution of the size indicates that the twisted CME expands slightly slower than the writhed one.

Overall, the evolution of these two properties (magnetic field and radial size) shows that: (i) both CME models show an evolution with distance that is consistent with past results, (ii) it is not possible to distinguish between the two models on the basis of their evolutionary signature alone, and, (iii) the twisted CME corresponds to a small CME with strong magnetic fields, whereas the writhed one corresponds to a large CME with weak magnetic fields. This result may be associated with the fact that the writhed CME is significantly faster than the twisted CME. 

\subsection{Evolution of the CME Speed}

There have been relatively fewer studies of the evolution of the CME speed, because statistical studies are not especially appropriate for these. The reason is that  CMEs have larger speed variation close to the Sun (ranging from 200 to 3000~km\,s$^{-1}$), but their magnetic field strength is probably more closely clustered within a factor of 4.  We determined the maximum speed along the ecliptic in the meridional plane through the CME nose. 

\begin{figure}[ht!]
\figurenum{6}
\plottwo{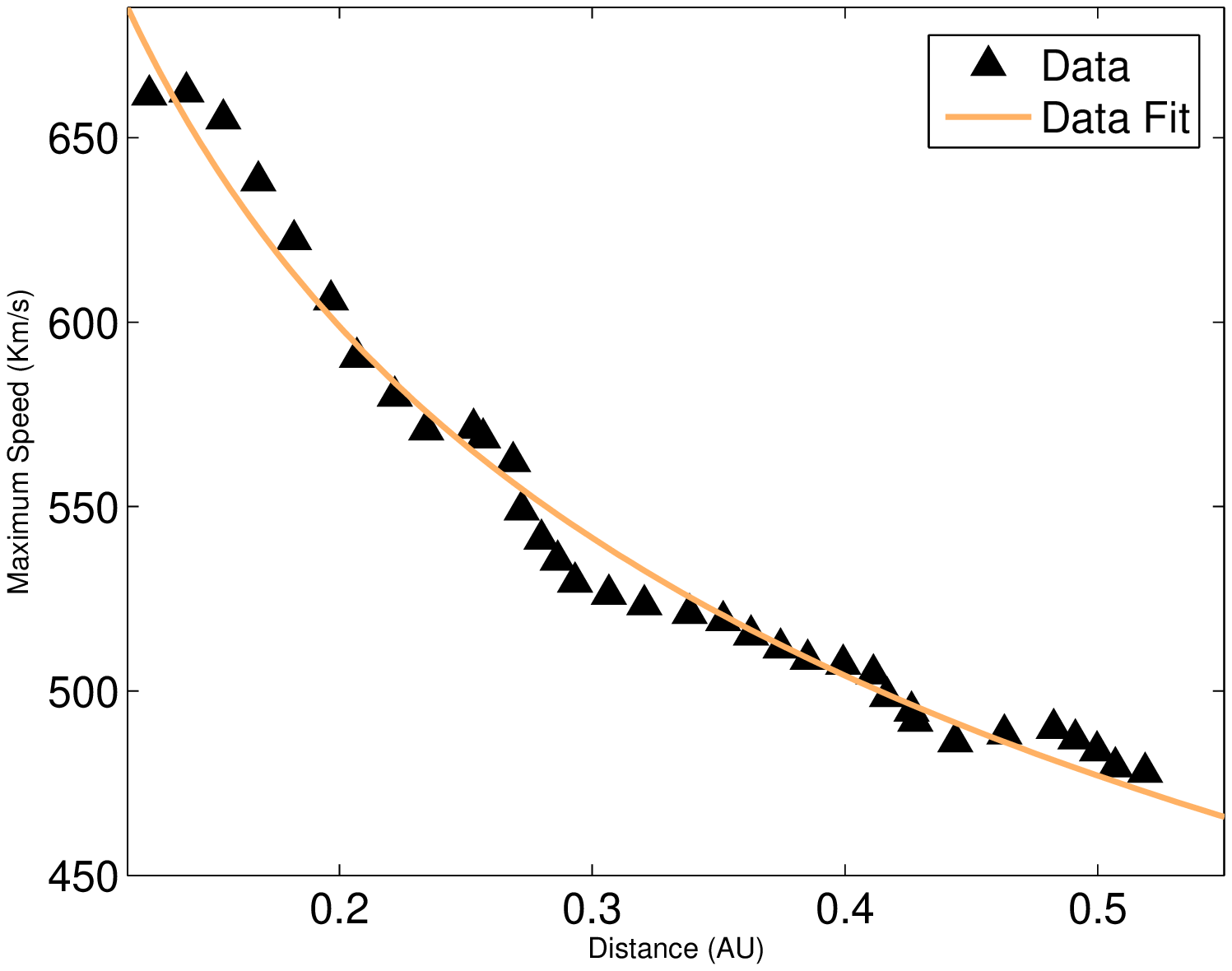}{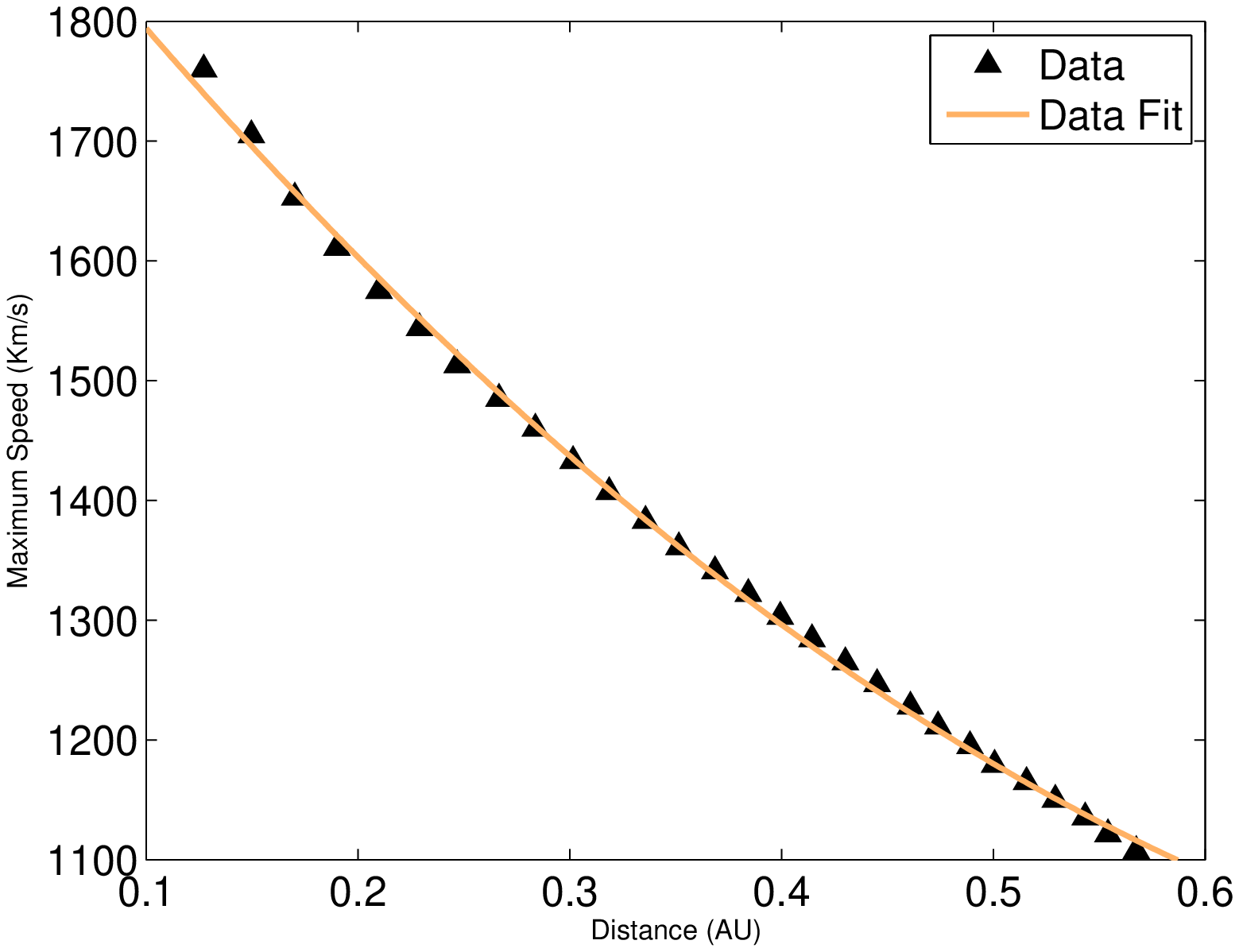}
\caption{Maximum speed of the CME for the twisted (left) and writhed (right) simulations as plotted versus the CME front radial distance. The orange lines show fits to the data: a power law for the twisted simulation, and a second-order polynomial for the writhed simulation.}
\label{fig:speed}
\end{figure}

Figure~\ref{fig:speed} shows the maximum speed of the CME (km~s$^{-1}$) versus the radial distance of the CME front (AU). We track the maximum of the speed within the boundaries of the magnetic ejecta (tracked to determine the radial size in the previous subsection), i.e.\ excluding the shock and sheath regions. Contrary to the evolution of the magnetic field and the size, the two simulations were best fitted with different functions.  The speed of the writhed CME  was better fitted with a second-order polynomial function rather than the power-law. In both cases, the CME decelerate as it propagates, which is expected for CMEs that travel faster than the solar wind. The fits are:
\begin{eqnarray*}
V_\mathrm{max, twist} &=&  401.7\, R^{-0.248},\\
V_\mathrm{max, writhe} &=& 2010.2 - 2285.2 \, R + 1252.7 \,R^2,  
\end{eqnarray*}
with an R-squared value of 0.981 and 0.9987, respectively.  As described above, it can clearly be seen that the twisted CME simulation corresponds to a relatively slow event (with a predicted speed at 1~AU of 402~km\,s$^{-1}$), whereas the writhed CME simulation corresponds to a much faster event (with an initial speed of 2010 km\,s$^{-1}$ and speed at 1 AU of 977~km\,s$^{-1}$). The power-law fit for the writhed CME simulation is $V_\mathrm{max, writhe} = 960 \, R^{-0.315}$, with a R-squared value of 0.978.

The study by \citet{Winslow:2015} found that the maximum speed of magnetic ejecta between 0.3 AU and 1 AU can be fitted as:
$$
V_\mathrm{max,Winslow} = 483\, R^{-0.26},
$$ 
which is somewhat similar to what we found for the twisted flux rope and the power-law fit for the writhed CME.

Most theories about CME deceleration rely on analogies with aerodynamical drag \citep[]{Gopalswamy:2001b, Cargill:2004,Vrsnak:2013}. In that sense, the CME decelerates due to a drag force associated with its interaction with the solar wind. This drag force is usually of the form $ - \rho\, A\, C_D\, (u-u_{sw}) |u-u_{sw}|$,where $u$ and $u_{sw}$ are the CME and solar wind velocity, respectively, $\rho$ is the solar wind mass density, $A$ is the CME cross-sectional area, and $C_D$ is the drag coefficient. 

As discussed in \citet{Cargill:2004} and \citet{Vrsnak:2013}, under typical circumstances, it can be assumed that $\gamma = \rho\, A\, C_D$ is constant with radial distance, as the area, $A$, approximately grows as $r^2$ and the density decreases as $1/r^2$. A further approximation is that the solar wind speed is nearly constant in the heliosphere (which is equivalent to assuming that the density decreases as $1/r^2$). Under these assumptions, the equation of the CME speed in the heliosphere, when the only force acting on the CME is the drag through its interaction with the solar wind, is:
$$ \frac{du}{dt} = -\gamma \, (u-u_{sw}) |u-u_{sw}|.$$

On the other hand, a power-law relation, as found here, is equivalent to $\ln{u} = \alpha \ln{r}$, with $\alpha$ being a constant. After differentiating, this gives:
$$\frac{du}{dt} = \alpha \frac{u^2}{r}.$$

Hence a power-law is equivalent to a drag-term where the drag coefficient decreases with distance as $1/r$. For the actual heliosphere, the interaction between CME and solar wind is not purely aerodynamical, and certainly involves magnetic forces. It is not out of question that the actual decrease of the drag term is stronger with distance than that considered by \citet{Vrsnak:2013}. The two simulations used in this paper appear to be consistent with this assumption. 

Heliospheric images of CMEs allow for the determination of the CME speed, albeit with significant uncertainties (due to the errors in CME tracking, but more importantly, due to the uncertainty in deriving radial distances from these observations). Work by \citet{Temmer:2011} and \citet{Liu:2013}, among others, have investigated the change of speed with time and/or distance, although they did not fit the velocity data to analytical models.

\section{Summary \& Conclusions} \label{sec:discussion}

In this article, we looked at the radial evolution of the properties of two simulated CMEs \citep[]{Jacobs:2009, Lugaz:2014}, which were compared with each other and with published results of past studies. We focused on those properties that are most often analyzed in statistical studies and remote observations: the CME radial size, the magnetic field strength inside the CME and the CME velocity. Comparing the twisted and writhed CME simulations with each other, it was found that the writhed CME is a faster, larger event with a speed that decreases faster with distance as compared to the twisted  CME event. Overall, in both simulations, the CME shows an expansion and a deceleration which agree well with results from previous studies, especially those focused on the inner-most heliosphere (between the Sun and 0.5~AU), i.e. those using MESSENGER in situ data, or remote-sensing observations. 

Theoretical work by \citet{Gulisano:2010} shows that the CME size should increase as $r^\zeta$, whereas the magnetic field should decrease as $r^{-2\zeta}$, with $\zeta = 0.7$. For each of the two simulations, we find that the index of the size increase is not a factor of two smaller than the index of the magnetic field strength decrease, as it would have been expected from that relation. Moreover, the size of the writhed CME increases slightly faster than that of the twisted one, but its magnetic field strength decreases slower (see further discussion on this theme in \citet{Dumbovic:2018}). 

We conclude that the simulations presented here are consistent with statistical results from previous studies as well as specific case studies. Furthermore, the radial trends of the simulated CMEs are nearly undistinguishable from each other. If we think of the heliospheric evolution of CMEs as being determined by the interaction of the CME with the solar wind plasma and IMF, it is somewhat expected that the evolution is primarily constrained by the change of the solar wind and IMF with distance. Following the launch of the Parker Solar Probe in August 2018 and the expected launch of Solar Orbiter \citep[]{Mueller:2013} in early 2020, we are expecting to have more frequent and better in situ measurements of CMEs in the innermost heliosphere. Our results can also be used to compare observations from these missions to the expected CME size and magnetic field strength at different heliocentric distances from two different numerical simulations.

\acknowledgments
This research was partially performed with funding from AGS-1433086. CJF and NL were partially supported by the STEREO grant 80NSSC17K0556. NL was also supported by NASA grants NNX15AB87G and 80NSSC19K0082.

%

\end{document}